# Energy Efficient Security Architecture For Wireless Bio-Medical Sensor Networks

Rajeswari Mukesh[1]
Dept of Computer Science & Engg
Easwari Engineering College
Chennai- 600 089

Dr.A.Damodaram[2]
Vice Principal
JNTU College of Engineering
Hyderabad-500 072

Dr.V.Subbiah Bharathi[3]
Dean Academics
DMI College of engineering
Chennai-601 302

*Abstract* **Latest developments in VLSI, wireless communications, and biomedical sensing devices allow very small, lightweight, low power, intelligent sensing devices called biosensors. A set of these devices can be integrated into a Wireless Biomedical Sensor Network (WBSN), a new breakthrough technology used in telemedicine for monitoring the physiological condition of an individual. The biosensor nodes in WBSN has got resource limitations in terms of battery lifetime, CPU processing capability, and memory capacity. Replacement or recharging of batteries on thousands of biosensor nodes is quiet difficult or too costly. So, a key challenge in wireless biomedical sensor networks is the reduction of energy and memory consumption. Considering, the sensitivity of information in WBSN, we must provide security and patient privacy, as it is an important issue in the design of such systems. Hence this paper proposes an energy efficient security protocol for WBSN where security is provided to the physiological data, which is being transmitted from the sensor node to the sink device. This is achieved by authenticating the data using patients biometric , encrypting the data using Quasi Group cryptography after compressing the image data using an energy efficient number theory based technique.**

## I. INTRODUCTION

The WBSNs [1,2] promise inexpensive [1,2], unobtrusive [1,2], and unsupervised ambulatory monitoring [1,2] during normal daily activities for prolonged periods of time. An example of WBSN is shown in Figure 1.To make this technology efficient and cheap, the tradeoffs between the system configuration and the security should be resolved. In WBSN the sink nodes collect data from the mobile patients through biosensors and is then transmitted to the healthcare provider for health monitoring[2]. In the existing system the authentication between biosensor nodes and sink devices, and between sink devices and healthcare providers is performed only by using Message Authentication Code (MAC). In the proposed system, security is provided in two steps. In the first step, the data that is being transmitted from the biosensor nodes to the sink device is compressed and encrypted using Quasigroup encryption algorithm to provide confidentiality. In the second step, the strong authentication of the biosensor nodes to the sink device is done by using the Heart rate variability of the patient. The beat-to-beat heart beat interval[3] is used as a biometric characteristic to generate the identity of the individual. This usage of biometrics allows automatic identification/ verification of individuals by their physiological characteristics. This type of authentication detects intruders entering into the sensor network between the sensor node and the sink device there by securing the data against meet-in-the-middle attack.

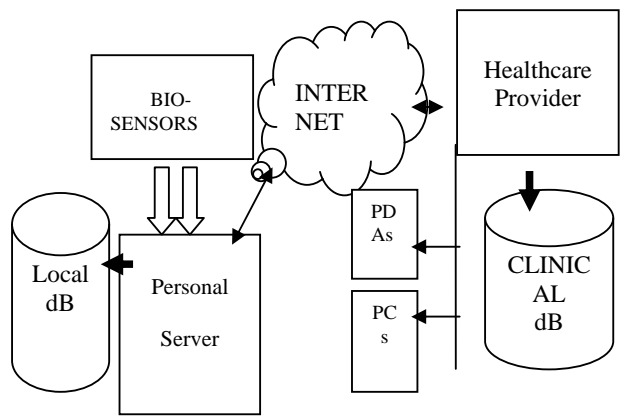

**Figure 1. Wireless Biomedical Sensor Network**

## II. RELATED WORK

Wireless Sensor Network was initially designed without taking the security aspect into consideration. Research has been carried out in the field of security protocols for WSN for providing authentication and confidentiality security service. Authentication using MAC and biometrics like ECG [3] and EEG [4] has been already proposed by many researchers in different levels. Shu-Di-Buo [5] et. all has proposed mutual biometric authentication using ECG or PPG between two sensor nodes that are intended to communicate with each other. The communication range between these sensor nodes is very short. Hence the possibility attack is also very less and this scheme does not provide any solution to overcome the attack between the sink device and the biosensor nodes.

For providing confidentiality security service the security architectures like, SNEP, Zigbee makes use of AES, RC5 kind of symmetric cipher. These symmetric cryptographic algorithms are complex and hence they are not energy efficient.

.

## III. OUR CONTRIBUTION

In the proposed system an energy efficient two level security architecture is designed for WBSN. In the first level, the physiological data taken from the individual is transmitted from the biosensor nodes to the sink device securely by compressing it using a number theory based CRT technique and then encrypting it using an energy efficient quasigroup encryption. The Chinese Remainder Theorem (CRT) [6] has been a useful tool in applications of number theory to other fields. The CRT is based on the solution of linear and modular congruencies. The congruence is nothing more than a statement about divisibility. Since it makes use of few arithmetic operations it is considered to be energy efficient to compress the biomedical images. Quasi groups [7] (or Latin squares) provide a powerful method for generating a larger set of permutation transformations by permuting not only the samples but also transforming the amplitudes themselves across their range. By doing this, they provide an immensely large number of keys, even for small alphabets. Therefore, quasi group based ciphers can have a variety of applications, and also strong in overcoming brute force attacks. It has been proved that the quasi group transformation maximizes the entropy at the output, which is desirable for a good system. This system provides extremely large group of keys that ensures enhanced security. It can work either in the chain mode or in the block mode. Block mode is more tolerant to errors compared to the chain mode. The following Table I [7] shows the no. of latin squares used in quasigroup encryption. It is clear from the Table I that if the value of n increases, the task of breaking the quasi group cipher is of astronomical complexity. Thus if the key is temporary, it would be very difficult to extract the information using brute force. The known plain text attack and replay attack are also not possible because the key keeps changing. Thus using energy efficient Quasigroup encryption provides the confidentiality security service.

Biometrics is a metric that is commonly used for automatic identification or verification of persons by his or her own taken from the body surface. Some of the famous biometrics are fingerprint, iris pattern and hand geometry etc, are patterns taken from the body surface. But the biometric used in WBSN is a physiological sign generated by a biological system of an individual like heart rate variability (HRV) or EEG signals.

Table I. Bounds on the number of Latin squares for n = 16, 32, 64

| N | Lower Limit (no. of Latin squares) | Upper Limit (no. of Latin squares) |
|---|---|---|
| 16 | $.101*10^{119}$ | $.689*10^{138}$ |
| 32 | $.414*10^{726}$ | $.985*10^{784}$ |
| 64 | $.133*10^{4008}$ | $.176*10^{4169}$ |

HRV is estimated by taking the inverse of the time interval between the peaks of adjacent R waves (RR interval) in ECG. HRV has been shown to be unique [3] for different subjects, which satisfy the basic criteria of a biometric characteristic. Physiological signals such as HRV are time variant which make them difficult to be applied in conventional biometric systems. In the proposed system two sensors placed at different locations of the individual, will capture their own copy of biometric characteristic independently but simultaneously at time t. The HRV of these two sensors are proved to be identical or highly correlated. If the two sensors were not on the same individual, i.e) if any one of the sensor nodes is compromised, HRV measured from the two sensors will not be identical. Thus the authentication of biosensor nodes by the sink device is done. This proposed architecture is intended to apply to telemedicine and related fields, where data collected by the sensors for medical application are now as well used intelligently as biometric characteristic for sink device to recognize the sensors placed on a human body.

## IV. OVERALL ARCHITECTURE

Wireless Sensor Network is becoming a promising technology for various applications. One of its potential deployments is in the form of Wireless Biomedical Sensor Network (WBSN) for measuring physiological signals. The architecture of secure WBSN is illustrated in Figure 2. The miniature wireless intelligent module, which can be integrated with some kind of biosensor, is referred as WBSN node. Physiological signals (EEG, ECG, Temperature, Blood pressure, Glucose level etc) measured by wearable or implantable biosensors are gathered by the sink device and transmitted to the healthcare provider via 3G network.

The server at the healthcare provider stores the data into patient database, do long term trend analysis and prediction. The data are published via web service. The healthcare professionals and patients can access the long term physiological data via internet. WBSN provide long term and continuous monitoring of patients under their natural physiological states even when they move.

Then clinicians can analyze the physiological data and give diagnosis advices accordingly. Alternatively, when a clinician is away from the hospital, he/she still can get the data via a PDA and give diagnosis advices to the patient remotely. This system provides convenience for patients as well as for clinicians. For patients, they can get medical service at home or any other places they prefer. And they can move around freely while carrying light hand-held medical device. For clinicians, they can give diagnosis suggestions to patients remotely without the necessity to go to the hospital if nothing emergency happens.

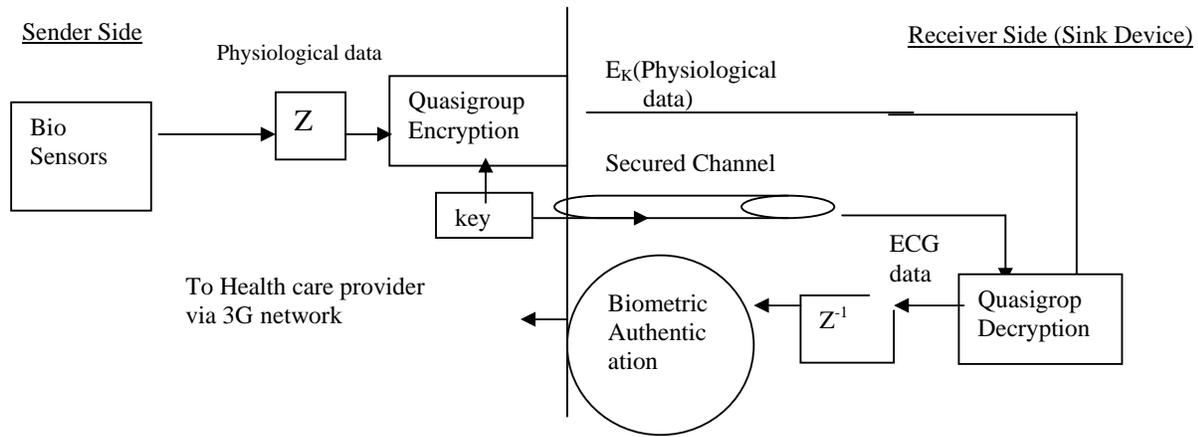

**Figure 2 : Overall Architecture**

The two level security for providing confidentiality and authentication security service for the physiological data transmitted from the sensor node to the sink device is clearly illustrated in Figure 2.

## V. DESIGN OVERVIEW

The following assumptions are made in this design
(i) Biosensors have acquired initial certificate from the trusted third party, got authenticated from the sink and has a shared key with which the ECG data has to be encrypted.
(ii) The routing functionalities are being taken care by the network layer and not included in this work.
(iii) Wearable biosensors are used for data acquisition.
(iv) Noise in the data is eliminated by means of filters.

### A. Compression Using Chinese Remainder Theorem

The Chinese Remainder Theorem (CRT) [6] has been a useful tool in applications of number theory to other fields. The CRT is based on the solution of linear and modular congruencies. The congruence is nothing more than a statement about divisibility.

Given a system of congruencies to different moduli:
$x \equiv a_1 \pmod{m_1}, x \equiv a_2 \pmod{m_2} ... x \equiv a_r \pmod{m_r}$ and if each pair of moduli are relatively prime, i.e.) $\gcd(m_i, m_j) = 1$ for $i \neq j$, the system has exactly one common solution modulo. $M = m_1 * m_2 * ... m_r$ and any two solutions are congruent to one another modulo $M$. The Chinese Remainder Theorem can be used to increase efficiency by making use of relatively small numbers in most of the calculation.

The merits of CRT are as follows that suits WBSN
- Increased efficiency in machine computation.
- Reduced memory, and sophisticated hardware requirements
- Reduction in space requirement for storage of data because large numbers are converted into relatively smaller ones by solution of linear congruencies.
- Use of simple arithmetic operations like addition, subtraction, multiplication, and division and hence execution of Million Instructions Per Second (MIPS) is possible.
- Faster computation process and hence reduction in processing time.
- Widespread application in cryptography, secure transmission of codes and signals in military and defense applications.

The algorithm for image compression using CRT is as follows.

The images are generally represented in the form of *NxM* matrix. In color image coding applications[6] the color spaces, namely red, green and blue in 24 bits per pixel (bpp) RGB scale of 8 bpp each are compressed separately as in the gray scale image. An image of size *NxM* is taken and is fragmented into blocks of size *1xK*. Each pixel r*[i]* in the block is divided by 16 to produce two half pixels of 4 bits each.

$$a[i] = r[i]/16, \ i = 1 \text{ to } k \quad (1)$$
$$a'[i] = r[i] \bmod 16, \ i = 1 \text{ to } k \quad (2)$$

Thus the input image is considered as a sequence of half pixels $a[1,2...k]$, $a'[1,2...k]$ and the key sequence is a set of relatively prime numbers given by
$$n[1,2...k] > a[i] \text{ and } a'[i]$$

Image: $a[1,2...k]$, $a'[1,2...k]$ - block of half pixels

Key: $n[1,2...k]$ -> set of relatively prime integers, can also be generated by a cryptographically strong random number generator such as BBS. Now generating *N* for each key value using *P*, where *P* is the product of all the keys, calculates the Coefficients of the CRT using equation 5.

.

$$N[i] = P / n[i] \text{ where } P = \prod n[i]. \quad (3)$$

using the equation 4 linear congruencies are generated

$$N[i] * x[i] = 1 \left( mod\ n[i] \right) \quad (4)$$

where $x[i]$ satisfies the above congruency and

$$C[i] = N[i] * x[i] \quad (5)$$

These stages are carried in prior to transmission, the values of $C[i]$ can be generated once the key is decided; hence they are calculated and stored in the system to be used during transmission. For the transmission of the image, the value of TR is determined for each block of $k$ half pixel values as follows.

$$TR = \sum C[i] * a[i] (mod P) \text{ - Cipher Text (quotient)} \quad (6)$$

$$TR' = \sum C[i] * a[i]' (mod P) \text{ - Cipher Text (remainder)} \quad (7)$$

For $k$ half pixel values, one $TR$ and $TR'$ value is transmitted providing compression; moreover, this value is dependent on the key used which incorporates encryption. This is the most vital step of the algorithms as it ensures simultaneous encryption and compression.

At the receiving end, the $k$ half pixel values are regenerated from the single value $TR$ and $TR'$.

$$ar[i] = TR \left( mod\ n[i] \right) \text{ -Plain Text quotient} \quad (8)$$

$$ar'[i] = TR' \left( mod\ n[i] \right) \text{ -Plain Text remainder} \quad (9)$$

The pixels are then reconstructed from the half pixels using equation 10.

$$s[i] = ar[i] * 16 + ar'[i] \quad (10)$$

As explained in the previous section, the encoded image, to be transmitted, is given by

In equation 6 $C[i]$ are pre-calculated coefficient and $a[i]$ are the pixel values after applying the threshold. Since $C[i]$ are pre-calculated, they need not be calculated for every $TR$. The reason for using Chinese Remainder Theorem[6] for solving the linear congruencies is to reduce a bigger number to a smaller representation. For image of size N x M and block size K, all (N x M)/K TR are computed. After computing all TR, the frequency of each distinct TR and their counts are determined. They are sorted in descending order of their count and assigned new set of numbers.

A table of unique TR and an equivalent code is generated. Using this table each TR obtained is encoded into this new code. The same is followed for TR'. At the receiver, the same encoding table is used to recalculate the actual TR and TR' values from which the half pixel values ar[i] and ar'[i] and thus the reconstructed image pixels s[i] are determined. Since the integrated encryption done is lacking security level, one more level of encryption is done by using Quasigroup[7] technique and is described in the next algorithm.

### B. Data Enciphering And Deciphering

In order to provide additional security to WBSN by considering the limitations in memory and processing capability, the information that is being transmitted between the sensor node/head and the base station can be encrypted using Quasigroup encryption algorithm. Quasigroups [7] are well suited for encrypting physiological signal related information. The strength of this encryption has been already examined[7]. A groupoid [8] is a finite set Q that is closed with respect to an operator *, i.e., a* b ∈ Q for all a, b ∈ Q. A groupoid is a quasigroup, if it has unique left and right inverses, i.e., for any u, v ∈ Q there exists unique x and y such that x*u = v and u * y = v. This means that all operations are invertible and have unique solutions, which implies their usability as cryptographic substitution operations. A quasigroup can be characterised with a structure called Latin square. A Latin square is an n x n matrix where each row and column is a permutation of elements of a set. In this case |Q| = n.

The requirements for WBSN information encryption are the following:
- It must be computationally easy for the biosensor node to encrypt the data using quasigroup encryption and send it to the sink device.
- It must be computationally easy for the sink device to decrypt the cipher text using Quasigroup decryption.
- Computationally faster in the network environments which has a limited processing power and other resource constraints.
- Be compact enough for use in sensor node memory space.

### C. Biometric Authentication between Biosensor Node and Sink

As described in section 3 the ECG signal at time t is taken from the two biosensors placed on the same individual and the difference between their HRV are calculated using Hamming distance. A normal ECG trace consists of a P wave, a QRS complex and a T wave. A small U wave may also be sometimes visible, but is neglected in this work for its inconsistency. An ECG sample is shown in figure 3. The P wave is the electrical signature of the current that causes atrial contraction; the QRS complex corresponds to the current that causes contraction of the left and right ventricles; the T wave represents the repolarization of the ventricles; and the U wave, although not always visible, is considered to be a representation of the papillary muscles or Purkinje fibers. The QRS complex is the most characteristic waveform of the signal with higher amplitudes.

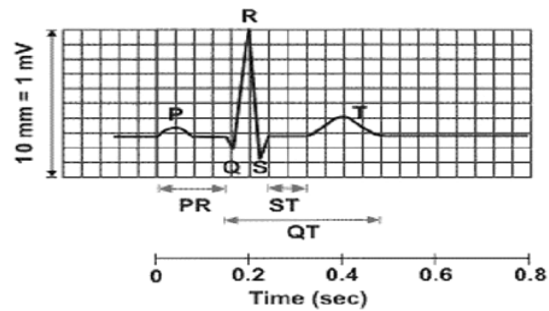

P wave (0.08 - 0.10 sec)
QRS ( 0.06 – 0.10 sec)
P-R interval (0.12 – 0.20 sec)
Q-T interval (0.2 - 0.4 sec)

**Figure 3: ECG Sample**

Using Hamming distance data comparison of HRVs of 2 sensor nodes are done. If the difference between HRV from two random sensor nodes exceeds a threshold, then an alarm signal will be raised to healthcare provider mentioning that the biosensor node has been compromised. The Hamming distance between two strings of bits (binary integers) is the number of corresponding bit positions that differ. This can be found by using XOR on corresponding bits or equivalently, by adding corresponding bits (base 2) without a carry. For example, in the two bit strings that follow:

```
       A     0 1 0 0 1 0 1 0 0 0
       B     1 0 1 1 0 1 0 1 0 1
A XOR B =  1 1 1 1 1 1 1 1 0 1
```

The Hamming distance ($H$) between these 10-bit strings is 9, the number of 1's in the XOR string. The following algorithm may be used to find the hamming distance.

```
Integer Hamdist(string value1, string value2)
Begin
        Integer dist=0
     Integer I=1
     Integer len
     If length(value1)>length(value2)
           then len=length(value1)
           else len=length(value2)

If((value1 is NULL) or (value2 is
                            NULL))
           Return NULL
     While(I<=len)
      dist=dist+(substring(value1,I,1)
         !=substring(value2,I,1))?1:0)
         I=I+1
     Return dist
End
```

## VI. RESULTS

### A. Compression using CRT

The biosensor image takes for analysis is MRI brain image shown in fig 4.

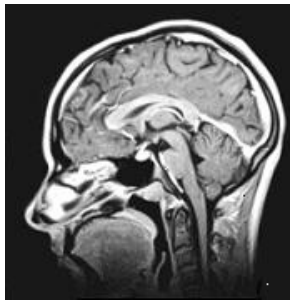

**Figure 4: MRI brain Image**

The implementation of the CRT compression algorithm is done in MATLAB 7 and the execution time and the compression ratio of the various compression algorithms are shown in table II.

.

**Table II. Comparison of compression ratio of various algorithms**

| Algorithm | Execution time in sec | Compression ratio |
|---|---|---|
| JPEG | 1.04 | 2.7603 |
| LZW | 0.65 | 6.2313 |
| SPIHT | 0.78 | 4.7714 |
| NTICE | 0.64 | 7.4057 |

### B  Quasigroup encryption

A part of the encoded MRI barin image shown in fig 4 is given in the screen shot figure 5 shown below and it is found that the quasigroup has more randomness in the cipher text than the complex AES symmetric cipher that has been proposed for WSN. The comparison of output randomness is shown in fig 6a & fig 6b.

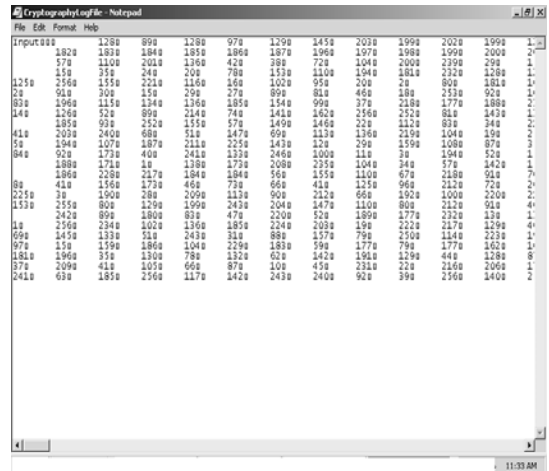

**Figure 5. Partial encoded compressed brain Image**

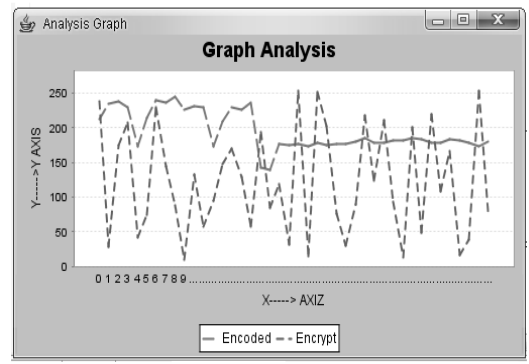

**Figure 6a: Graph   Analysis   For
    QuasiGroup Cryptography**

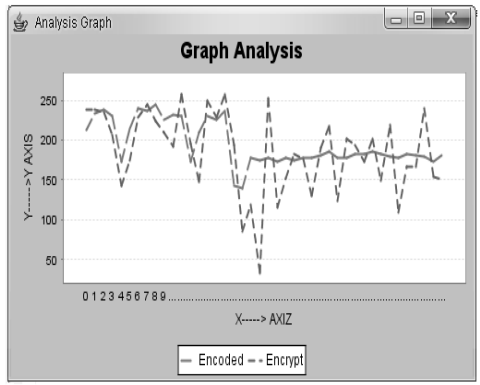

Figure 6b: Graph Analysis For AES Cipher

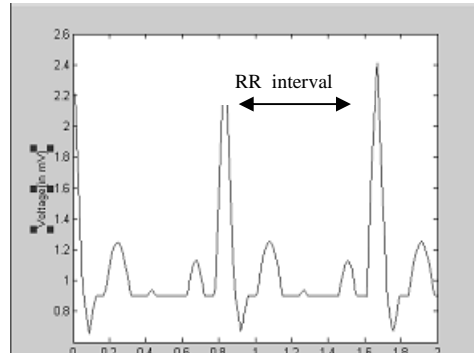

Fig 7b. ECG wave of sensor2 of person1

*C. Biometric authentication using HRV*

The R peaks have the largest amplitudes among all the waves, making them the easiest to detect and good reference points for future detections. The R peaks occur after the P peaks within 0.1 seconds. The R peak occurs between 0.12 seconds and 0.2 seconds after the P peak. Q and S peaks occurs about the R peaks within 0.1 seconds. The QT interval lies within 0.44 seconds. The MATLAB output for R peak detection for two ECG sensors placed on the same individual with a heartbeat rate of 72 is shown in Figure 7. The RR interval for both the ECG waves are 0.84 sec and HRV is 1.19. Hence the difference is zero and they are proved to be from the same person. The MATLAB output for R peak detection for two ECG sensors placed on two different persons with a heartbeat rate of 72 and 60 respectively is shown in Figure 8. The RR interval for the ECG wave of sensor1 from person1 is 0.84 sec and HRV is 1.19, and the RR interval for the ECG wave of sensor2 of person2 is 1and HRV is 1. The difference between HRV is 0.19 and hence this indicates that one of the node has been compromised by an intruder.

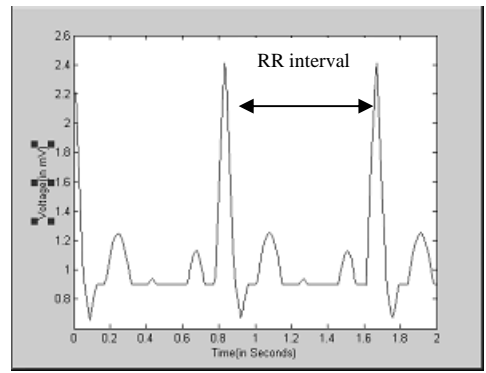

Fig 8a. ECG wave of sensor3 of person1

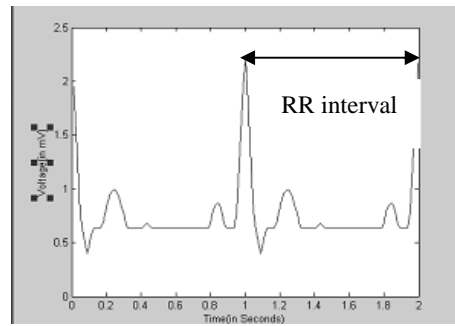

Fig 8b. ECG wave of sensor2 of person2

## VII. CONCLUSIONS AND FUTURE WORK

The proposed scheme detects malicious biosensor nodes using biometric authentication and provides confidentiality using Quasigroup encryption after compressing the biomedical images using CRT.These techniques are proved to be energy efficient which is an important requirement of WSN. Among the patient's vital signals, ECG generates the highest data rate which is about 10 kB/s. R-interval analysis can be performed to determine the peaks. By transmitting R-intervals instead of the whole ECG waveform, the data rate can be lowered and power consumption can be reduced subsequently. ECG signals are easy to identify and relatively immune to potential noise interferences. The future work is to build experiment environment based on the proposed

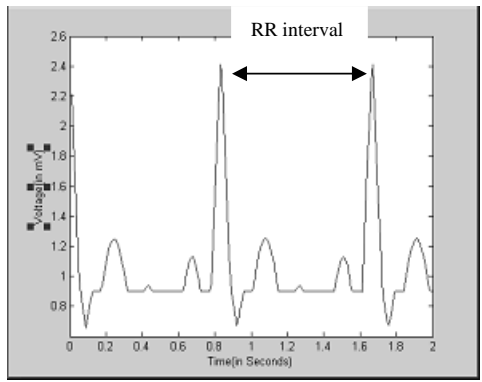

Fig 7a. ECG wave of sensor1 of person1

.

system and to extend the security at the health care provider level.

**REFERENCES**

[1] DDKouvatsos, G Min and B Qureshi. Performance issues in a Secure Health Monitoring Wireless Sensor Networks, *Performance Modeling and Engineering Group, University of Bradford*, Bradford BD7 1DP, UK, WP.

[2] Aleksandar Milenkovic, Chris Otto, Emil Jovanov, Wireless Sensor Network for Personal Health Monitoring : Issues and an Implementation, *Electrical and Computer Engineering Department, The University of Alabama, Huntsville*, AL 35899.

[3] L. Biel, O. Pettersson, L.Philipson and P.Wide, "ECG Analysis: A New Approach in Human Identification," *IEEE Trans on Instrumentation and Measurement*, vol.50, pp.808-812, June 2001.

[4] Sebastein Marcel, Jose del R.Millan, Person Authentication Using Brainwaves (EEG) and Maximum A Posteriori Model Adaptation, *IEEE Conference on pattern analysis and machine intelligence*, April 2007, Vol.29, No.4, pp 743-752.

[5] Shu-Di-Bao, Yuang-Ting Zhang, Liang-Feng Shen, Physiological Signal based Entity Authentication for Body Area Sensor networks and Mobile Healthcare systems, *IEEE Conference proceedings of Annual International Conference of the IEEE Engineering in Medicine and Biology Society*, 2005,3:2455-2458.

[6] Vikram Jagannathan, Aparna Mahadevan, Hariharan and Srinivasan, "Number Theory Based Image compression Encryption and Application to Image Multiplexing", © 2007 *IEEE - ICSCN*, Feb. 2007, pp.59-64.

[7] Maruti Venkat, Kartik Satti, A Quasigroup Based Cryptographic System, *International Journal of Network Security,* July 2006, Vol.7, No.1, pp. 15–24.

[8] Marko Hasinen, Smile Markovski, "Secure SMS messaging using Quasigroup encryption and Java SMS API"

[1]Rajeswari Mukesh is working as Assistant Professor in Department of Computer Science and Engineering at Easwari Engineering College, Chennai. She has received her B.E and M.E in Computer Science and Engineering and currently pursuing Ph.D at JNTU Hyderabad. Her area of interests include Network Security and Image Processing.

[2]Dr. A. Damodaram is Director, ASC and Professor of Computer Science & Engineering, JNTU College of Engineering, Hyderabad. His research interests include Software Engineering, Computer Networks and Image Processing. Prof. Damodaram was awarded his Ph.D. in Computer Science & Engineering from JNTU . He has a rich experience of 17 years in Teaching, Research and mentoring research scholars in his respective areas. He is Member of Academic Council in Cochin University of Science and Technology, Cochin. He is a member of AIEEE, New Delhi and Governing Council, JNTU College of Engineering, Hyderabad.

[3]Dr.V.Subbiah Bharathi is working as Dean Academics at DMI Engineering College, Chennai. He has received Ph.D from Manonmaniam Sundaranar University. He has got 25 national and international papers published in reputed journals including ACM. His area of research include Image processing and Network Security .

.